\documentclass{ws-p8-50x6-00}

\begin{document}

\title{Recent Results from PHENIX}

\author{Kenneth N. Barish for the PHENIX Collaboration}

\address{University of California at Riverside,
Riverside, CA 92506, USA \\ 
E-mail: Kenneth.Barish@ucr.edu} 

\maketitle

\abstracts{I report on results from the PHENIX collaboration for the
  first RHIC run, where gold ions were collided at $\sqrt{s_{NN}}=130~GeV$. 
  In order to study initial conditions, PHENIX has measured the ratio
  between anti-protons and protons, the transverse energy, and charged
  multiplicity as a function of collision centrality. These
  measurements allows us to extract some information on
  the initial density of created partons, the baryon density, the
  energy density, and particle production. In order to probe the
  formation of a possible new state of matter, PHENIX has measured the
  neutral pion and unidentified hadron transverse momentum spectra.
  We find a deficit compared to the the expectations from simple
  nucleon-nucleon scaling. These novel results are consistent with
  expectations that high $p_T$ leading particles will be suppressed in
  the presence of the formation of a plasma.
 }

\section{Introduction}

Can we create matter in which quarks are free? What makes up the
majority of the mass around us? What makes up the spin of the proton? 
These are some of the questions that the PHENIX collaboration, which 
utilizes the Relativistic Heavy Ion Collider (RHIC) at Brookhaven National 
Laboratory, will address.
RHIC is a versatile collider which can produce Au+Au collisions up to
200~GeV/u, polarized p+p collisions up to 500~GeV, as well as ion
combinations in between. During the first RHIC run
during the summer of 2000, gold ions were collided at 130~GeV/u  from
which the PHENIX detector recorded 5 million events.

The PHENIX detector is designed to measure a broad range of signals
including rare penetrating probes, such as photons, electrons, and
muons, that are sensitive to the formation of a deconfined state of
matter. In order to accomplish this, PHENIX is a complex integrated
detector system that has a high rate capability, high granularity,
good mass resolution, and good particle ID. The central detector was
approximately 50\% instrumented for the summer 2000 experimental
run.\cite{bib:zajc}

In this paper I report on a subset of PHENIX's results from the
first year of data taking. I subdivide these results into two
categories: initial conditions and hard probes. The results are
intriguing in their own right and show the promise of the program 
over the coming years.

\section{Initial Conditions}

Initial conditions are required to characterize the state of nuclear 
matter formed. To fully understand the initial conditions
a comprehensive program that includes e+A, p+A, and a variety of A+A
species are needed. However, even with A+A collisions at full energy we
can begin to get a handle on some of the initial conditions.

From the first year of data of Au+Au collisions at full energy,
PHENIX has measured the ratio between anti-protons and protons, the 
transverse energy, and charged multiplicity as a function of collision 
centrality. This allows us to extract some information on the 
initial density of created partons, the baryon density, the energy 
density, and particle production.

\subsection{Gluon saturation}

Gluons can begin to fuse with high enough gluon density, limiting the
parton production. If the gluon density is indeed high enough to
saturate, it implies that the system would thermalize quickly. This would
be important as it has implications on the initial energy density achieved.

\begin{figure}[t]
\centering
\epsfxsize=20pc 
\epsfbox{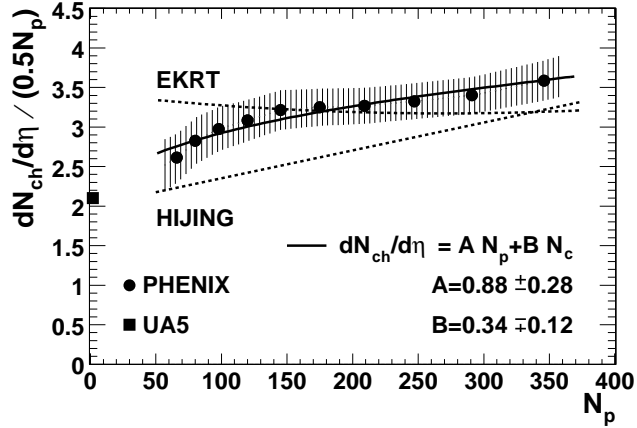} 
\caption{Charged-particle pseudorapidity density per participant pair
  vs. the number of participants.\protect\cite{bib:mult}\label{fig:mult}}
\end{figure}

If the parton production is limited, we expect that the final state
charged particle yields to also be limited. Therefore,  
PHENIX's measurement\cite{bib:mult} of the charged particle multiplicity as a
function of number of participants constrains models. (See
Fig.~\ref{fig:mult}). In particular,
the results disfavor models that have a fixed saturation
scale,\cite{bib:eskola} while models in which the saturation scale
changes with centrality can describe the
data.\cite{bib:karzeev} While the latter class of models can describe the
data, it is not a unique description. Further measurements with
different species and energies, and perhaps ultimately e+A collisions, are
needed to explore whether the initial gluon density saturates.

\subsection{Baryon density}

The measure of the baryon and energy density is required to map out the phase
diagram of nuclear matter. Lattice calculations, which predict that
nuclear matter undergoes a phase transition, are calculated at zero
baryon density. 

In a preliminary analysis, PHENIX has measured $\bar{p}/p = 0.64\pm
0.01\pm 0.08$. This represents a net baryon density significantly
lower than at the SPS,
and is approaching the low baryon density limit where lattice
calculations are preformed. The ratio of $\bar{p}/p$ shows only a weak
centrality dependence and little or no $p_T$.

\subsection{Energy density}

Lattice calculations indicate the phase transition to occur at zero
baryon density and an energy density
$\epsilon_{crit}\approx 0.6-1.8~GeV/fm^3$. Assuming
a boost invariant expanding cylinder of dense nuclear matter 
in which a thermalized system is reached after a formation time, 
$\tau_0$, Bjorken derived a formula from which the measured transverse
energy, $E_t$ can be related to the energy density, $\epsilon_{Bj}$.
For the most central events, we measure\cite{bib:et} $\epsilon_{Bj}=4.6~
GeV/fm^3$ for a formation time of 1~fm/c. This is roughly
1.5 to 2 times higher than previous experiments. However, we expect
the formation time to be smaller at
RHIC than at the SPS, perhaps indicating a much higher
energy density at RHIC. For example, if the formation time is 0.2~$fm$
we arrive at an energy density of $23~GeV/fm^3$.
Therefore, it appears that the the energy deposition is certainly
adequate, so the
question becomes whether a thermalized system is created fast enough
before $\epsilon$ falls below $\epsilon_{crit}$. The PHENIX program includes
the measurement of direct $\gamma$'s, which should help conclude
whether we are indeed producing a thermalized system.

\subsection{Particle production}

Many models of particle production identify two components: (A) soft
interactions where production scales with the number of participating
nucleons; and (B) hard interactions where production scales with the
number of binary collisions. Therefore, we write
$dN_{ch}/d\eta = A\times N_{part}+B\times N_{bin}$.
PHENIX has measured\cite{bib:mult} $B/A=0.38\pm 0.19$ from our charged particle
multiplicity measurement. (See
Fig.~\ref{fig:mult}). We also find that the transverse
energy density per participant scales in a consistent manner with the
charged particle density per participant.\cite{bib:et} These results
provide evidence for a term in the growth which scales like the number
of collisions.

\section{Hard Probes}

It has been predicted\cite{bib:quench} that partons respond
differently to a deconfined plasma
phase compared to ordinary nuclear matter, therefore providing a probe
of the formation of a plasma. Parton-parton interactions at high $Q^2$ 
occur during the very early pre-equilibrium phase of nucleon-nucleon
collisions. At a high enough $Q^2$, pQCD can be used to reliably
calculate hard scattering rates. Using these rates and measured parton
distribution and fragmentation functions, the rates of high $p_T$
leading hadrons can be predicted.

Partons are expected to lose energy via gluon radiation in traversing
a quark-gluon plasma. The partons then fragment into the hadrons which
we measure. Above 2 GeV we expect the hadrons to be largely a result of 
jet fragmentation. Therefore, we look for a suppression of high $p_T$ leading 
particles, which carry information about whether the momentum of the 
parton is lowered while traversing the formed matter.

\subsection{PHENIX's $\pi^0$ and unidentified leading hadron measurements}

\begin{figure}[t]
\centering
\epsfxsize=30pc 
\epsfbox{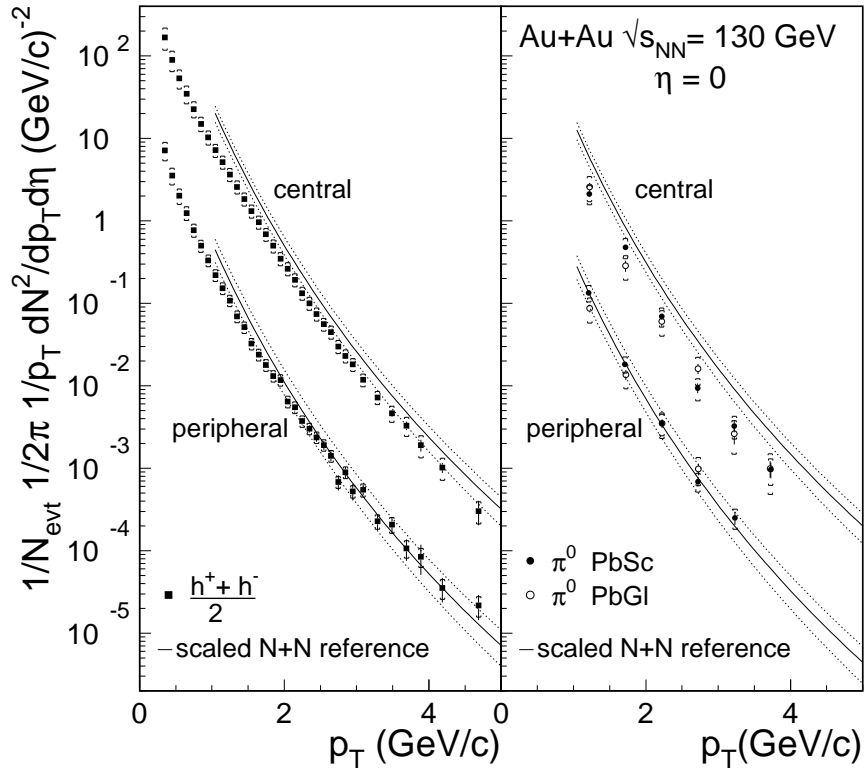} 
\caption{The yields per event at mid-rapidity for (left) charged
  hadrons and (right) neutral pions, with the $\pi^0$ results from the
  PbSc and PbGl analyses plotted
  separately.\protect\cite{bib:highpt}\label{fig:highpt}}
\end{figure}

PHENIX has measured\cite{bib:highpt} the transverse momentum spectra of $\pi^0$
and unidentified charged hadrons out to $\approx$~4~GeV from Year-1 of
RHIC. Currently, there
is no N+N data at 130~GeV to serve as a baseline, and therefore we
parameterize the p+p and p+$\bar{p}$ charged hadron spectra and
interpolate to 130~GeV. We then scale by the number of binary
collisions in two different centrality classes: ``peripheral''
representing 60-80\% and ``central'' representing 0-10\%. For the
$\pi^0$ baseline we make
the additional correction due to the $h/\pi$ ratio measured at the ISR.
When we compare our measured leading particle spectra with the p+p baseline we
find that for the peripheral centrality class it is consistent with
N+N scaled by the number of collisions, while for the central
centrality class it is significantly below the scaled N+N spectra. We
also observe that the $\pi^0$ deficit is larger than that from the
unidentified hadrons. (See Fig.~\ref{fig:highpt}.)
This is consistent with our identified spectra
measurements which show that the particle composition is a strong
function of $p_T$. 

\subsection{Complications}

So far I have ignored possible complications due to nuclear
effects. However, the transverse momentum spectra is known to 
be modified due the ``Cronin effect'' and quark and gluon ``shadowing''.

The Cronin effect can be modeled as prior parton scattering. This has
the effect of broadening the transverse momentum spectrum, which {\bf
enhances} high $p_t$ particles. Calculations show that the largest
enhancement is expected around $p_T$~=~4~GeV/c, and to become
small above 6~GeV/c.\cite{bib:wang1}

Nucleon structure functions are known to be modified in nuclei,
translating into a {\bf deficit} of high $p_T$ particles. In the
frame where the nucleon is moving fast, this can be modeled as a
recombination effect due to the high gluon number density at low x. The
nuclear shadowing due to quarks has been measured. For the x range
relevant to RHIC, $x>10^{-2}$, it is around 10\% effect. Unlike
quark shadowing, gluon shadowing has not been measured. Furthermore,
gluon scatter will be important at RHIC. However, in the x range
relevant for RHIC is expected to be smaller than a 10\%
effect.\cite{bib:sarcevic}

Therefore, the current understanding of the known nuclear effects cannot
account for our experimental observation. Our central collisions data
shows significant suppression relative to the prediction without
energy loss. This indicates a novel effect: the deviation from point
like scaling. It is consistent with parton energy loss, but without p+p
data at the appropriate energy, p+A data to verify the expectations for
quark and gluon shadowing, and a larger $p_T$ reach to eliminate the
possibility contributions from collective soft effects, it is too early
to make definitive conclusions. 

\section{Outlook}

The first PHENIX run was a huge success. In addition to commissioning
a complex detector, our charged particle multiplicity measurements
constrain models, the energy density appears to be well above the phase
transition level, and we have suggestive high transverse momentum
measurements.

The second PHENIX run is ongoing. During this run we expect RHIC to
achieve design luminosity. Assuming RHIC delivers the promised
integrated luminosity, PHENIX will be able to measure high-$p_t$
leading particles out beyond 10~GeV, and make  measurements of
the $J/\Psi$, $\omega$, $\phi$, and direct-$\gamma$'s. Equally as
exciting, the Year-2 will include the first polarized proton
collisions.


\end{document}